\documentstyle[12pt]{article}
\title{Causal, psychological, and electrodynamic \newline 
  time arrows as consequences of the \newline 
  thermodynamic time arrow}
\author{Hrvoje Nikoli\'c  \\
Theoretical Physics Division, Rudjer Bo\v{s}kovi\'{c} Institute, \\
P.O.B. 1016, HR-10001 Zagreb, Croatia \\
{\normalsize hrvoje@faust.irb.hr} \\
\makebox[1in]{} \\
}
\date{\today}
\begin{document}
\maketitle
\begin{abstract}
 A clear explanation is given on how the causal, psychological, and
electrodynamic time arrows emerge from the thermodynamic time arrow. 
\end{abstract}

\mbox{ }\newline
Most of physicists agree that the causal, psychological, and electrodynamic
time-arrows are consequences of the thermodynamic time-arrow. 
For example, such a viewpoint is  
explicitly expressed in famous books \cite{feyn}, 
\cite{hawk1} (see also \cite{leb} for some other references). However, 
it is our feeling that the existing literature does not  
explain the relations among various manifestations of 
the time arrow in a sufficiently clear way, and that there is still a lot of 
misunderstanding among some physicists about the true origin of the 
time arrow. We believe that we are able to clear up how the 
causal, psychological, and electrodynamic time arrow emerge from the 
classical, deterministic physical laws and the assumption that 
disorder increases with time.       
 
Let us start with the causality principle. By causality we understand 
a ``fact"   
that ``causes" happen before ``consequences". Such a
claim does not make any sense at the fundamental level because 
the state of a physical system at any instant $t$ can be viewed as a 
``cause" of states for all other instants, larger 
or smaller than $t$, in the
sense that it uniquely determines these states via the equations of   
motion. However, it does make sense at some effective, macroscopic level.
This is connected with the fact that, in practice, we try to draw    
conclusions about the future and about the past without knowing 
{\em details} of the present state. Thus, our conclusions are to a
great extent based on statistical arguments. And since disorder 
increases with time, statistical arguments can determine the past
much better 
than the future. Thus it is easy to see a connection between
the presence and the past, but it is not so easy to see a connection
between
the presence and the future. That is why we can consider the past as a
``cause" of the presence, whereas can not consider the future as a
``cause" of the presence.

This can also explain the psychological time-arrow. Our psychological
feeling that the time lapses is a consequence of the fact that we
remember, i.e., our brain possesses the information about the events 
that do not refere to the present time. The problem of the origin
of the psychological time-arrow is reduced to the question     
why we remember the past, but not the future. 

The general mechanism  
of memorizing is the following: Some event in the past causes some
permanent consequence. Recalling is just observation of this consequence at
an arbitrary later instant. In other words, by observing the present
state of a system, we may draw the conclusion about its past, i.e., about the
event that caused the present permanent consequence.  
As we have already discussed, we cannot so easily draw   
conclusions about its future, except that the permanent consequence will
keep its form yet for some time, which is not the information about
any {\em event} in the future.   

Let us now discuss the status of causality in classical
electrodynamics (similar arguments can also be applied to all other 
classical field theories). We consider a system which consists of   
charged currents and electromagnetic fields, both being time dependent.
We usually assume that the vector potential $A^{\mu}(x)$ is given
by the retarded solution of the equation
\begin{equation}\label{19}
 \Box A^{\mu}(x) = J^{\mu}(x) \; . 
\end{equation}
It is usually argued that we have to take the retarded solution because
of the causality principle, i.e., because the cause (moving charges) must
precede the consequence (electromagnetic fields). However, the essential 
tacit assumption is that the charges {\em are} the cause and that the
electromagnetic fields {\em are} the consequence. Such a viewpoint is 
partly influenced by equation (\ref{19}), where it is tacitly 
assumed that $J^{\mu}(x)$ is some known function, whereas $A^{\mu}(x)$  
is a function to be determined from (\ref{19}). However, this is
only an idealization. Electrodynamics is actually a system of
equations, one of which is (\ref{19}), whereas the other is the differential
equation for charged ``matter" fields, interacting with  
electromagnetic fields. This system 
should actually be solved self-consistently,
so there is no separation of fields into ``causes" 
and ``consequences" at the fundamental level. The electromagnetic field
is the cause of motion of the charge equally as the motion 
of the charge is the cause of the electromagnetic field. 

Furthermore, advanced solutions are not unphysical at
all. For example, it is possible to construct a large spherical source of
electromagnetic waves which emits waves from the sphere to the inner
area. When these waves come to the center, they force the positive and
negative charges in the center to oscillate. In the inner area of the sphere
this looks just as the advanced solution of (\ref{19}). 

However, there is
one peculiarity with this picture. One could argue that this would 
be inconsistent
if positive and negative charges were not present in the center at the
beginning. However, it is clear that there is no incosistency from the point of
view of quantum electrodynamics; the strong fields in the center can create
electron-positron pairs. This is also consistent with the classical 
{\em field} theory of charged matter, because nothing prevents induction of
local charge densities as long as the total charge is conserved.     
     
Although the advanced solutions are not unphysical, 
it is still true that the retarded solution is the appropriate one
for most of practical purposes. This is because in practice it  
is rather difficult to prepare such initial conditions (such as a large spherical
source) that would correspond to the advanced solution in a large volume
$V$. This can also be explained by the principle of disorder increase;
in a typical physical situation, the energy is first concentrated inside the
small lumps of matter (small disorder) and then dissipated all around via
electromagnetic waves (large disorder).

Finally, let us stress that the fact that disorder is growing with time is
equivalent 
to the statement that the Universe was quite ordered in the past. Thus, 
the only real problem with the time arrow is to explain {\em why} 
the Universe was
so ordered at some instant of time of its evolution. There is still no 
convincing explanation of this, except the anthropic principle
\cite{hawk2}.

\section*{Acknowledgement}

This work was supported by the Ministry of Science and Technology of the
Republic of Croatia under Contract No. 00980102.

\end{document}